\documentclass[fleqn,usenatbib]{mnras}

\usepackage{newtxtext,newtxmath}

\usepackage[T1]{fontenc}

\DeclareRobustCommand{\VAN}[3]{#2}
\let\VANthebibliography\thebibliography
\def\thebibliography{\DeclareRobustCommand{\VAN}[3]{##3}\VANthebibliography}

\usepackage{graphicx}	
\usepackage{amsmath}	

\title[Impact of hydro solvers on cloud-wind interactions]{Sensitivity of non-radiative cloud-wind interactions to the hydrodynamics solver}

\author[J.R. Braspenning et al.]{
Joey Braspenning,$^{1}$\thanks{E-mail: braspenning@strw.leidenuniv.nl}
Joop Schaye,$^{1}$
Josh Borrow$^{2}$
and Matthieu Schaller$^{3,1}$
\\
$^{1}$Leiden Observatory, Leiden University, PO Box 9513, NL-2300 RA Leiden, The Netherlands\\
$^{2}$Department of Physics, Kavli Institute for Astrophysics and Space Research, Massachusetts Institute of Technology, Cambridge, MA 02139, USA\\
$^{3}$Lorentz Institute for Theoretical Physics, Leiden University, PO Box 9506, NL-2300 RA Leiden, the Netherlands
}

\date{Accepted 2023 April 20. Received 2023 April 15; in original form 2022 March 25}

\pubyear{2023}

\begin{document}
\label{firstpage}
\pagerange{\pageref{firstpage}--\pageref{lastpage}}
\maketitle

\begin{abstract}
Cloud-wind interactions are common in the interstellar and circumgalactic media. Many studies have used simulations of such interactions to investigate the effect of particular physical processes, but the impact of the choice of hydrodynamics solver has largely been overlooked. Here we study the cloud-wind interaction, also known as the ``blob test'', using seven different hydrodynamics solvers: Three flavours of SPH, a moving mesh, adaptive mesh refinement and two meshless schemes. The evolution of masses in dense gas and intermediate-temperature gas, as well as the covering fraction of intermediate-temperature gas, are systematically compared for initial density contrasts of 10 and 100, and five numerical resolutions. To isolate the differences due to the hydrodynamics solvers, we use idealised non-radiative simulations without physical conduction. We find large differences between these methods. SPH methods show slower dispersal of the cloud, particularly for the higher density contrast, but faster convergence, especially for the lower density contrast. Predictions for the intermediate-temperature gas differ particularly strongly, also between non-SPH codes, and converge most slowly. We conclude that the hydrodynamical interaction between a dense cloud and a supersonic wind remains an unsolved problem. Studies aiming to understand the physics or observational signatures of cloud-wind interactions should test the robustness of their results by comparing different hydrodynamics solvers.

\end{abstract}




\section{Introduction}
Interactions between cold, dense clouds moving at high velocity through a hot and dilute medium are common in turbulent layers within the circumgalactic medium (CGM) \citep[][]{tumlinson_circumgalactic_2017}. Such layers are hypothesized to originate in cold, dense clouds finding themselves entrained in a hot, dilute flow. Such a configuration could, for example, arise from a hot, feedback-driven wind interacting with pre-existing or newly formed clouds, or if gas clumps fall through a hot, hydrostatic halo. The turbulent, intermediate-temperature gas created when the cloud is stripped may find itself, under the right conditions, near the peak of the cooling curve and can condense rapidly to the cold phase. This process allows for the cloud's total cold gas mass to grow \citep[e.g.][]{gronke_how_2020, gronke_growth_2018}.

The precise fractions of cold and intermediate-temperature gas formed in the cloud-wind interactions are crucial towards making observational predictions \citep[e.g.][]{shelton_modeling_2012, de_la_cruz_simulated_2020}. The robustness of these predictions is not aided by the theoretical uncertainty about whether clouds fragment into a mist \citep{mccourt_characteristic_2018} or whether a few large clouds contribute the bulk of gas flux \citep[e.g.][]{gronke_is_2020, vijayan_kinematics_2020}.

In the past few years a tremendous amount of work has gone into improving the physical models included in simulations of cloud-wind interactions to study the effects of magnetic fields \citep[e.g.][]{sparre_interaction_2020, sparre_physics_2019}, conduction \citep[e.g]{kooij_efficiency_2021, armillotta_survival_2017}, rich multiple cloud systems \citep[e.g.][]{banda-barragan_shock-multicloud_2020, banda-barragan_shock-multicloud_2021} and cosmic ray driven winds \citep[e.g.][]{bruggen_launching_2020}. The fractions of cold gas as a function of time differ with the specific choice of model, but generalised analytical considerations gave rise to criteria governing the long-term survival of the cold cloud \citep[e.g.][]{li_survival_2020, kanjilal_growth_2021}. An intuitively reasonable conclusion arose: if the gas cools faster than it gets disrupted, the cloud will survive. While different authors would agree on this general statement, the precise cloud survival criteria they obtain differ, and are in some cases mutually exclusive. We will argue that many of the differences could actually originate from the different hydrodynamics solvers used. 

When solving a hydrodynamics problem without an analytical solution, such as the cloud-wind problem (also called the ``blob test'' in the hydrodynamics literature), astrophysicists have to rely on numerical simulations. Though these simulations should, in principle, all solve the same hydrodynamics equations, their methods differ widely. The underlying assumption when comparing relatively small, idealized simulations such as those of the cloud-wind problem, is that the methods will produce the same result given a sufficiently high resolution. This expectation is reasonable, since common methods are all rigorously tested and are known to be well behaved.

However, the cloud-wind interaction poses some unique challenges which can give rise to different problems for different methods. For example, pressure-confined clouds with high density contrasts are particularly challenging for smooth particle hydrodynamics (SPH) methods, whereas the symmetry inherent to the problem may pose a problem for grid methods. Furthermore, SPH and grid codes tend to respectively under- and overestimate the mixing rate of different gas phases \citep[][]{hopkins_new_2015}.

Notable work has been done specifically on the comparison between SPH and grid methods \citep[e.g.][]{agertz_fundamental_2007}. Motivated by those findings, many improvements have been made to SPH methods to reduce spurious surface tension using artificial diffusion and conduction. As the cloud-wind interaction involves many phenomena for which no analytical solution is available, there has been a considerable effort to deconstruct it into simpler, better understood constituent pieces. For example, extensive comparisons for the Kelvin-Helmholtz (KH) instability test using well-posed initial conditions have shown that, when using higher order kernels, SPH, grid and spectral methods agree both in their resulting morphology and quantitative evolution \citep[e.g.][]{tricco_kelvinhelmholtz_2019, lecoanet_validated_2016, mcnally_well-posed_2012}. 

Moreover, new hydrodynamics methods have been adopted with the aim of bridging the gap between a static Eulerian grid and Lagrangian SPH. Codes that have become widely used in astrophysics are \textsc{Arepo} \citep{springel_e_2010} and \textsc{gizmo} \citep{hopkins_new_2015}. \textsc{Arepo} uses Voronoi tiling to create cells around quasi-Lagrangian cell generating points, with the goal of combining the flux exchange of grid methods with the conservative properties of SPH methods. The two \textsc{gizmo} methods, meshless finite mass (\textsc{mfm}) and meshless finite volume (\textsc{mfv}), in effect build on the Voronoi principle by introducing smoothed boundaries between neighbouring cells, much like in SPH \citep[][]{vila_particle_1999}. The \textsc{mfm} flavour does this by conserving the mass of volume elements at the time of flux exchange, whereas the \textsc{mfv} methods conserves their volume. 

Rapid development in the SPH community has also led to a large number of new flavours, all using slightly different diffusion and viscosity prescriptions in order to perform well on a suite of hydrodynamics tests. However, even though modern hydrodynamics methods are converging on a standard set of analytically solvable test problems, the cloud-wind interaction has only rarely been compared beyond approximate morphological similarity. Despite the growing literature using the cloud-wind interaction to model additional physics and make observational predictions, the uncertainties in the non-radiative hydrodynamic result have received little attention.

To investigate the uncertainty in the results of simulations of the cloud-wind problem, and give quantitative measures of their magnitude, both in terms of theoretical quantities and observational characteristics, we carefully study seven different hydrodynamics methods with the same initial conditions.
These seven cover all methods used in existing studies into the wind-cloud problem we are aware of and should hence provide a comprehensive picture of the differences and similarities. They are:
\begin{enumerate}
 \item Modern Smooth Particle Hydrodynamics (SPH; three flavours)
 \item SPH-ALE (two flavours: \textsc{mfm}, \textsc{mfv})
 \item Moving Mesh (\textsc{mm})
 \item Static grid with adaptive mesh refinement (\textsc{amr})
\end{enumerate}
Each of these methods has been used in large cosmological simulations and extensively tested on a range of hydrodynamics test cases. All have advantages and disadvantages, which are especially apparent in difficult problems such as those with large amounts of rotation or vorticity. 

Because the interaction between a dilute wind and a dense cloud has no analytic solution, there is no independent benchmark to compare the results with. Instead, parallels to other, simpler, problems are drawn and signs of such behaviours are sought. For example, one looks for Kelvin-Helmholtz instabilities at the edges of the cloud, even though the cloud is curved and being pushed along which makes the comparison difficult.

Furthermore, code presentation papers for the above methods are often inconsistent in their choice for density thresholds, colour scales, and definitions of shown quantities, making like-for-like comparison neigh impossible. This work will homogenise such comparisons and add new quantities which shed new light on similarities and differences between the seven methods.

Here we use non-radiative hydrodynamics simulations to isolate differences due to the hydrodynamics solvers. As our simulations are non-radiative and do not include physical conduction, they should, after the initial phase in which viscosity can generate entropy in shocks, and in the limit of infinite resolution, only show evolution in the gas densities and temperatures due to the pressure gradients induced by the ram pressure exerted by the wind. Non-adiabatic energy exchange between different phases only takes place at the microscopic level, which is not modelled by any of these methods. We stress that this is a scenario never achieved in nature, where some mixing is always expected to occur. Some of the methods included in this work model mixing either implicitly or numerically, but none of them include a physical model based on the microscopic exchange of energy, which gives rise to the emergent macroscopic, finite volume, mixing.

Volume elements in finite resolution simulations represent averages over many microscopic phases, which can artificially give the appearance of a single mixed phase. Such numerical mixing is inherent to grid and mesh methods, but also occurs in SPH methods where artificial conduction is invoked to suppress spurious surface tension. It is often implicitly assumed that the level of mixing predicted by reasonably converged high-resolution simulations is realistic. However, this may be optimistic given that the mixing is still numerical as long as microscopic conduction is not modelled.

An interesting new avenue that we will explore, is to compare the level of mixing in a lower resolution simulation to that in a higher resolution simulation after coarse-graining the latter onto a lower-resolution grid for analysis. The hope is then that as one increases the resolution, the coarse-grained results converge even if the highest resolution simulation is still not converged. This would indicate convergence of the volume-averaged phase structure. However, we note that observational diagnostics such as emission and absorption signatures depend on the microscopic rather than the coarse-grained phase structure.

We study the level of deviation between these different methods and find that it is large enough to have a major impact on predictions for observations, as well as on the theoretical quantities typically measured in cloud-wind interactions.

In section \ref{sec:methods} we introduce the hydrodynamics methods and the set-up used in this study. In section \ref{sec:results} we then compare the simulations using different methods using a number of different metrics. First we compare the evolution of dense mass in section \ref{sec:results-dense_mass}, then we look at the amount of intermediate-temperature gas in section \ref{sec:results-mass_intermediate_temp}, we consider what covering fraction this gas produces in projection in section \ref{sec:results-covering_fraction}, and finally study the effect of coarse-graining in sections \ref{sec:downsampling}. We conclude and make some recommendations in section \ref{sec:conclusions}.

\begin{figure*}
    \centering
    \includegraphics[width=\textwidth]{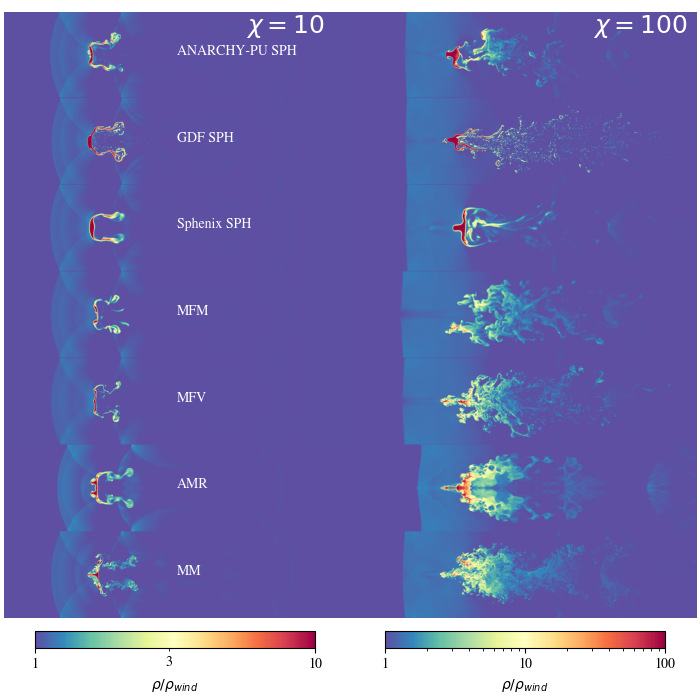}
    \caption{Density slices of the simulation box at $t \approx 5t_{\text{cc}}$, for initial density contrasts $\chi = 10$ (left) and $\chi = 100$ (right), at the highest resolution ($n=256^3$). Large morphological differences can be observed.}
    \label{fig:slice_plots}
\end{figure*}

\section{Methods} \label{sec:methods}
In this work we simulate cloud-wind interactions using seven different hydrodynamics solvers. These solvers are selected to represent a range of Lagrangian and Eulerian algorithms which have been used in cosmological simulations and/or idealised experiments. All simulations are non-radiative and use identical initial conditions in which the cold cloud and hot wind are in thermal pressure balance. This relates their temperatures and densities as
\begin{equation}
    \frac{T_{\text{wind}}}{T_{\text{cloud}}} = \frac{\rho_{\text{cloud}}}{\rho_{\text{wind}}} \equiv \chi \, .
\end{equation}
where $\chi$ is the density contrast.
The wind is set to have a Mach number $\mathcal{M} = \frac{v_{\text{wind}}}{c_{\text{s,wind}}} = 1.5$, with $v_{\text{wind}}$ the wind speed and $c_{\text{s,wind}}$ the wind's sound speed. The wind and the cloud are initialised on a grid with respectively \emph{N} and $\chi^{1/3}$\emph{N} resolution elements on a side. The region into which the cloud is inserted is then removed from the wind's grid. Our simulations are wind tunnels that are four times longer than they are wide and high. All boundary conditions are periodic, a choice that we found to have no significant effect on the results presented here. In our default set-up the cloud has a radius of one-tenth of the box width. As we use pure hydrodynamics, i.e. radiative cooling is disabled, the problem is scale invariant. In the unit system used throughout we choose
\begin{equation}
    R_{\text{cloud}} = 0.1 \, .
\end{equation}
Two important time scales used throughout the results section are the cloud crushing time and the drag time.
The cloud crushing time
\begin{equation}
    t_{\text{cc}} = \frac{\sqrt{\chi} R_{\text{cloud}}}{v_{\text{wind}}} \, ,
\end{equation}
with $v_{\text{wind}}$ the wind speed, is the time it would take an external shock hitting the cloud's edge to travel to the cloud centre, taking into account the lower sound speed by a factor of $\sqrt{\chi}$ inside the cloud due to its higher density. Though the cloud is not expected to have fully disappeared by this time, it is a useful measure. In practice, clouds survive for $\approx$10 $t_{\text{cc}}$.

The drag time
\begin{equation}
    t_{\text{drag}} = \frac{\chi R_{\text{cloud}}}{v_{\text{wind}}} \, ,
\end{equation}
is the time it would take a cubical cloud to sweep up its own mass in wind particles. For a spherical cloud this time is larger by $\frac{4}{3}$, which is commonly ignored.

In this study we use three modern versions of SPH:
\begin{enumerate}
    \item The new density-energy implementation of \textsc{Sphenix sph} \citep{borrow_sphenix_2022} integrated in the SWIFT simulation code \citep{schaller_swift_2016, schaller_swift_2018} , which was particularly designed to work well with sub-grid physics modules for galaxy formation and includes limiters to achieve that.
    \item The SPH method \textsc{gdf sph} (Geometric Density Force) as implemented in SWIFT. This is an independent implementation of the equations presented in the \textsc{gasoline-2} methods paper \citep{wadsley_gasoline2_2017}, using the same parameter values as used by those authors.
    \item The \textsc{anarchy-pu sph} pressure-energy method as implemented in SWIFT, which is based on \citet{schaller_eagle_2015} (see also appendix A of \cite{schaye_eagle_2015}), but using energy instead of entropy as its thermodynamic variable \citep[][]{hopkins_general_2013}.
\end{enumerate}
The \textsc{mfm} (meshless finite mass) and \textsc{mfv} (meshless finite volume) methods are also used as implemented in SWIFT as the SPH-ALE method \citep{vila_particle_1999,lanson_renormalized_2008}, based on the equations presented in the \textsc{gizmo} methods paper \citep[][]{hopkins_new_2015}. 
The \textsc{gdf sph}, \textsc{anarchy-pu sph}, \textsc{mfm} and \textsc{mfv} methods are all independent re-implementations based on what the respective authors presented in their original methods papers.
The \textsc{mm} (moving mesh) method in \textsc{Arepo} \citep{springel_e_2010} and the \textsc{amr} (static grid with adaptive mesh refinement) method in \textsc{Athena}++ \citep{stone_Athena_2020} complete our set of seven hydrodynamics solvers.

The \textsc{mfm}, \textsc{mfv} and \textsc{amr} methods all use the HLLC Riemann solver, while the \textsc{mm} method uses an exact Riemann solver. The \textsc{amr} method uses a second order Piecewise Linear Method (PLM) reconstruction of the primitive variables, and the second-order accurate van Leer predictor-corrector scheme.
To ensure results independent of the chosen smoothing kernel, all SPH flavours use the same kernel (Quintic spline (M6)) and the same number of neighbours ($\eta = 1.23$, 83.49 $N_{\text{ngb}}$) (see \citet[][]{dehnen_improving_2012}).

There are three major differentiators between these three SPH methods. The first is the equation of motion that they choose to solve. \textsc{Sphenix sph} and \textsc{anarchy-pu} use different equations of motion derived from the same Lagrangian \citep[][]{hopkins_general_2013}, with different choices for the smoothed field. \textsc{gdf sph} explicitly symmetrises the equation of motion to remove spurious forces generated at contact discontinuities, yielding yet another equation of motion. 

The choice of the smoothed field, and hence the way the density is calculated, is the second differentiator. For \textsc{Sphenix sph} the density is computed as a smoothed quantity as a sum over neighbours. Instead, for \textsc{anarchy-pu sph} the pressure is smoothed. In comparison to the \textsc{Sphenix sph} equation of motion, this smoothed pressure formulation ensures that at points where there is an instantaneous jump in the internal energy (i.e. at the edge of the cloud) the pressure forces vary smoothly. If there is an associated increase in density, such that the pressure is intended to be uniform, then a smoothed density SPH can produce spurious repulsive forces (also known as surface tension) \citep[e.g.][]{hu_sphgal_2014}. In contrast, the smoothed pressure formulation does not create these as it ensures that information is used from both inside and outside of the cloud to construct the pressure field. Although this can lead to unintended time-stepping issues \citep{borrow_inconsistencies_2021}, it is highly effective at removing these instabilities.

The approach taken in \textsc{gdf sph} is to modify the equation of motion from the smoothed density formulation to reduce the impact of contact discontinuities. Their new equation of motion has been shown to reduce surface tension significantly \citep[][Fig. 1]{wadsley_gasoline2_2017}, but is no longer explicitly derived from a Lagrangian.

The third differentiator is the choice for artificial conduction and viscosity. Several prescriptions are present in the literature, and each of these methods chooses a different combination which they find to work best with their implementation.

In this work we choose to use the diffusion prescription and parameters as the respective original authors presented. We note that there is considerable disagreement in the literature which prescription performs best, and what scenarios each is applicable to. For example, \textsc{gdf sph} uses a model intended to mimic physical turbulent diffusion by following the trace of the shear tensor, whereas \textsc{Sphenix sph} uses a model aimed to limit the entropy generation in disordered fields and at contact discontinuities.


Throughout this work, all codes are run using their fiducial time-stepping, and where applicable the same CFL number has been used.

We have striven to make all initial conditions as similar as possible between all methods. All methods that are integrated in SWIFT use the same initial condition files, and the relevant parts of the initial condition generation code for \textsc{Arepo} are directly copied from the SWIFT equivalent. Because of the spherical shape of our cloud, there are some small and unavoidable differences at the interface of the cloud between the \textsc{amr} method and all others. In the \textsc{amr} method we allow two levels of adaptive mesh refinement based on the mass in each cell. This introduces some amount of splitting and merging of resolution elements that is not possible in any of the other methods. However, the existence of adaptive smoothing lengths and cell sizes should make such differences small. In the \textsc{mm} method, we also allow cell refinement and de-refinement based on cell mass, but in practice this is only relevant at very early times on the cloud boundaries. The density contrast of the cloud in the \textsc{mm} method is generated by having a larger number of cells, rather than by changing the mass in each cell. This is entirely analogous to the approach in the \textsc{sph}, \textsc{mfm} and \textsc{mfv} methods.

We model cloud-wind interactions with two different initial density contrasts each run at four different resolutions. Both these density contrasts, $\chi = 10$ and $\chi = 100$, may be high compared to those produced by realistic physical processes for clouds with sharp edges, but they are representative of the literature. We indicate the resolution by the number of wind particles or volume elements ($n$) per short side length of the simulation volume. In the elongated direction, four times that number is present. This brings the total number of wind particles or volume elements to $4n^3$. In the case of non-particle based methods, the initial centres of volume elements or mesh generating points are placed at the same locations as the particles in SPH. We make use of the resolutions $n = \{16, 32, 64, 128, 256\}$. Given the cloud radius, this translates to $nR_{\text{cloud}} = \{1.6, 3.2, 6.4, 12.8, 25.6\}$ wind particles per cloud radius\footnote{It is not clear that choosing a fixed number of particles or volume elements results in the same effective resolution. However, the resolution steps are the same for all methods, and hence the relative differences when changing the resolution are not affected by this uncertainty.}.

\section{Results} \label{sec:results}
Figure \ref{fig:slice_plots} shows the disruption of the cloud in progress for the seven different hydrodynamics solvers at the highest resolution. The figure shows infinitesimally thin slices along the mid-plane of the simulation volume\footnote{To achieve this we use the slice methods in the \textsc{swiftsimio} and \textsc{yt} python libraries, both use the gather approach to make slices. For the \textsc{amr} method a slice represents the density in the central cell.}. The lower density contrast ($\chi = 10$) is shown in the left column and the higher density contrast ($\chi = 100$) in the right column. Large differences between the different methods can be readily observed. All images are made at time $t \approx 5 t_{\text{cc}}$. It is clear that the problem is not scale free with respect to $\chi$. 

Noticeable is the offset of the bow shock in the \textsc{amr} simulation compared to the other methods. The shock observed in these images is formed from a conical bow shock which get wrapped onto itself by the periodic boundary condition. We confirmed that this wrapping around does not affect the quantities studied in this work. We speculate that the offset in shock position in \textsc{amr} is due to that method not being able to fully capture the diagonal movement of the conical bow shock, resulting in slightly different velocities and hence final position of the observed shock.

We observe that the morphological evolution of the \textsc{gdf sph} cloud seems different from that presented in \citet[][]{wadsley_gasoline2_2017}. However, those authors impose a much faster wind ($\mathcal{M} = 2.7$), causing ram pressure to dominate, and use a kernel (Wendland $\mathrm{C^4}$) with more than double the number of neighbours. A further difference is the limited range of our colour scale, which spans a factor of 10 in density for the initial density contrast $\chi = 10$, whereas those authors use a colour scale spanning a factor of 200. As noted before, the \textsc{swift} implementation of \textsc{gdf sph} is also completely independent from the code used by those authors. All of these factors make it hard to do a direct comparison. Remaining differences might be driven by slightly different initial conditions (see Appendix \ref{sec:ic_comparison}), different choices in the visualisation, or other yet unidentified factors.

Our results for \textsc{mfm} and \textsc{mfv} are smoother, though otherwise similar in shape to those presented in \citet[][]{hopkins_new_2015} (who also use $\mathcal{M} = 2.7$), especially the vertical fragmentation in the \textsc{mfv} method is reproduced. However, due to the use of a different Mach number those results are not directly comparable. \citet[][]{borrow_sphenix_2022} present a cloud exposed to a $\mathcal{M} = 2.7$ wind. Compared with our results, they find a qualitatively similar evolution with a central cloud holding together and only limited fragmentation or tail formation. Though all these code presentation papers used the higher Mach number $\mathcal{M} = 2.7$, we choose to adopt a lower value of $\mathcal{M} = 1.5$ in line with recent work exploring cloud-wind interactions in more detail \citep[e.g.][]{mccourt_characteristic_2018,gronke_growth_2018, gronke_how_2020, sparre_physics_2019, sparre_interaction_2020, kanjilal_growth_2021}.

We also note that at later times, when the cloud has largely dispersed in all schemes, the morphologies become very similar, except for the three SPH schemes when simulating the initial higher density contrast (see Fig. \ref{fig:projection_plots_late}). Furthermore, some authors choose to show the evolution of the entropy instead of the density, this makes it hard to directly compare different works as the entropy can have quite a different morphology from the density (see Fig. \ref{fig:projection_plots_entropy}). 
\subsection{Mass of dense gas} \label{sec:results-dense_mass}

\begin{figure}
    \centering
    \includegraphics[width = \columnwidth]{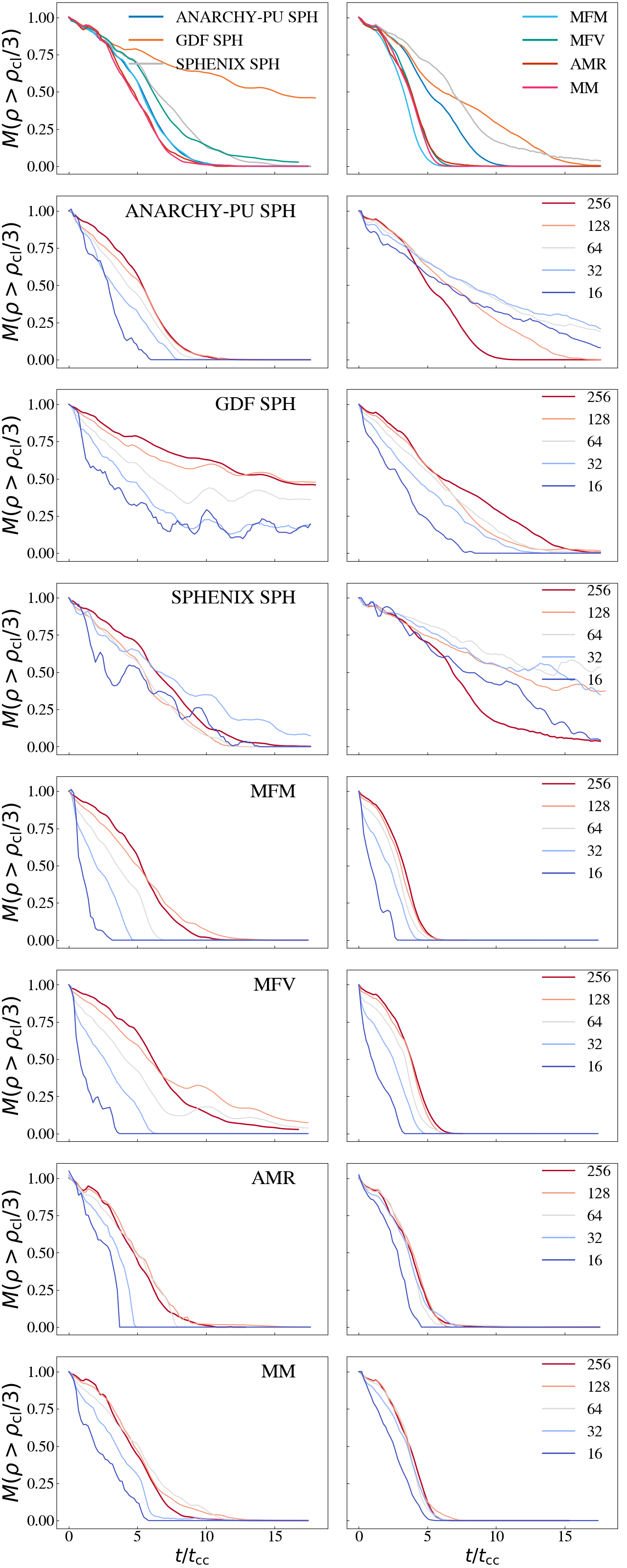}
    \caption{Evolution of the mass of dense ($\rho > \rho_{\text{cloud}}/3$) gas for different hydro schemes (colours and rows) and different resolutions (linestyles). Initial density contrasts are $\chi= 10$ (\textbf{left}) and $\chi=100$ (\textbf{right})}
    \label{fig:dense_mass}
\end{figure}

\begin{figure}
    \centering
    \includegraphics[width = \columnwidth]{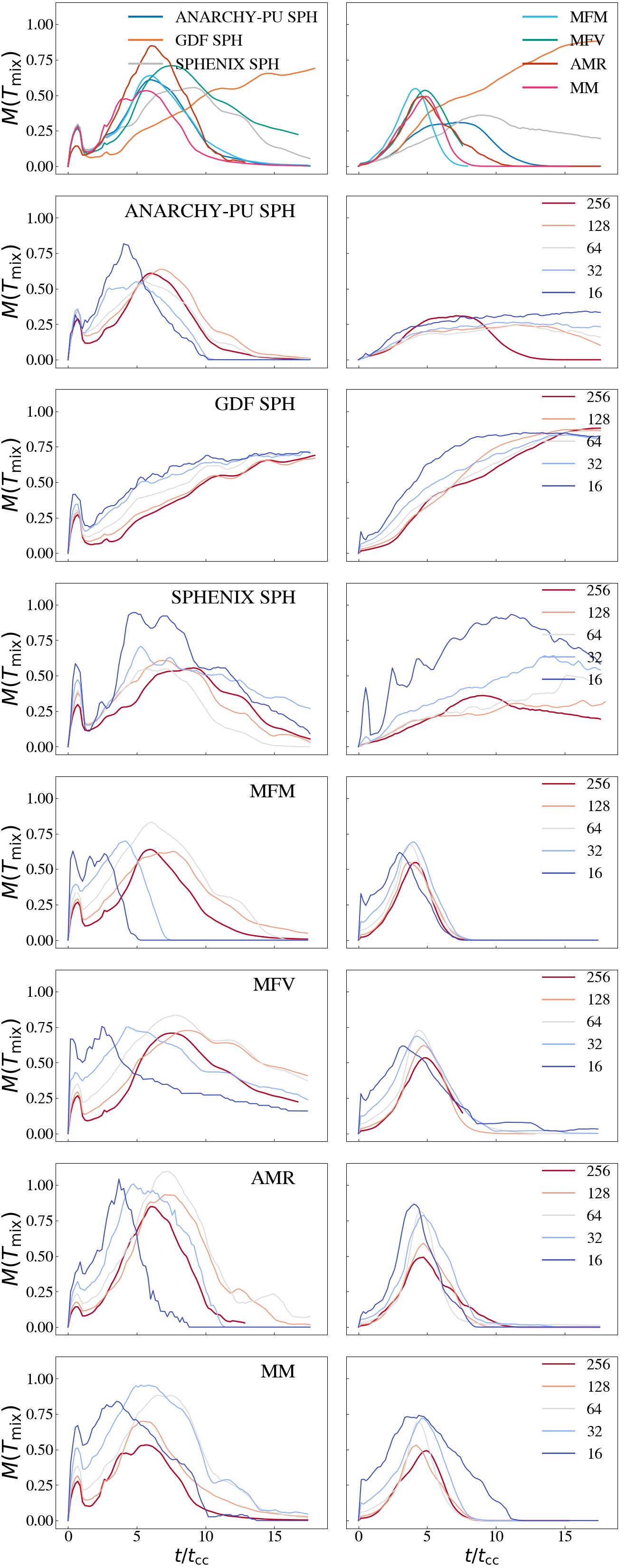}
    \caption{Evolution of the mass of intermediate-temperature gas for different hydro schemes (colours and rows) and different resolutions (linestyles). Initial density contrasts are $\chi= 10$ (\textbf{left}) and $\chi=100$ (\textbf{right})}
    \label{fig:intermediate_temp}
\end{figure}

The mass of dense gas in the simulation volume is defined as 
\begin{equation}
    M_{\text{dense}} = \sum_i m_i \left[\rho_i > \frac{\rho_{\text{cloud}}}{3} \right]\, ,
\end{equation}
where $\rho_{\text{cloud}}$ is the initial density of the cloud, $\rho_i$ the density of a particle or volume element and $m_i$ its mass. 
Figure \ref{fig:dense_mass} shows large differences in the evolution of the mass of dense gas between the different hydrodynamics methods for the two different initial density contrasts. The left and right columns show respectively the lower ($\chi = 10$) and higher ($\chi = 100$) density contrasts. The top row compares all hydrodynamics methods at the highest resolution ($n = 256$). Each of the remaining rows compares one method at all five resolutions. We remind the reader that even though none of the simulations explicitly model physical conduction, the high Reynolds numbers typical of astrophysical flows imply that turbulent conduction will become efficient below some scale. As mentioned above, a turbulent conduction model is included in \textsc{gdf sph}. At very high resolution, the effect of numerical mixing (i.e. mixing at the resolution limit) in mesh codes (\textsc{mm}, \textsc{amr}) is expected to be qualitatively and quantitatively similar to that of the mixing expected from turbulent conductions. At low or intermediate resolution the end result might be similar (i.e. everything is mixed), but the manner and time in which the cloud is destroyed may depend on resolution. However, because of the ram pressure and the initial shock front from the wind, pressure equilibrium is violated and, after an initial entropy generating phase, at later times the density may change adiabatically.

It immediately stands out that for the lower density contrast the cold cloud disperses much more slowly for the simulation using \textsc{gdf sph} than for the other methods. The results for simulations with the \textsc{anarchy-pu sph}, \textsc{mfm}, \textsc{amr} and \textsc{mm} methods are all very similar, while \textsc{Sphenix} \textsc{sph} and \textsc{mfv} are intermediate cases. The \textsc{mfv} method gives a slower decline in cloud mass, especially when the cloud has already lost more than 50\% of its original mass. Strikingly, this much slower dispersal at late times only occurs for the three highest resolutions. We note that for the \textsc{gdf sph} method the morphology of the dense gas is slightly clumpier (see Fig. \ref{fig:slice_plots}), with the clumps seemingly holding together indefinitely, becoming co-moving with the wind. This effect is amplified by increasing the resolution, suggesting that more of the gas is retained in dense clumps when the resolution is increased, though for the lower density contrast the two highest resolutions are similar in terms of the dense gas evolution.

Interestingly, for the lower density contrast, all methods seem quite well converged between the two highest resolution steps. At the higher density contrast, the convergence is distinctly worse for the SPH schemes, in line with the expectation that the higher contrast is a more challenging problem to solve. \textsc{Sphenix sph} and \textsc{anarchy-pu sph} show a sharp drop in the amount of dense mass at the highest resolution, more similar to the non-SPH methods, whereas \textsc{gdf sph} show the opposite trend with the decline becoming shallower, making it less similar at late times. The four non-SPH methods show excellent convergence between the highest three resolutions.

For the higher density contrast it is apparent that simulations with the three SPH methods have a much slower dispersal of the cold cloud. Whereas for the other four methods the cloud is completely destroyed within $\approx 7$ cloud crushing times, by that time the cloud (remnants) in the simulations using SPH methods still retain 30-60\% of their original mass. Though results from the other four methods seem similar when compared to SPH, the differences are non-negligible. The \textsc{mfm} simulation disperses the cloud the quickest, having lost half the cloud mass 25\% faster than the \textsc{amr} simulation, which has the longest cloud survival. The \textsc{mfv} and \textsc{mm} simulations are very well matched, which is expected given that the \textsc{mfv} method should be similar to the \textsc{mm} method bar having smoothed cell boundaries. However, this level of similarity contrasts with the large difference at late times between the \textsc{mfv} and \textsc{mm} simulations for the lower density contrast. We note that for the lower density contrast the threshold for dense gas is much closer to the wind density, making it more likely that fluctuations in the background density contribute to the mass of the dense gas.

It is well known that \textsc{amr} methods suffer from numerical mixing due to gas being exchanged across cell boundaries and mixed at the resolution scale \citep[e.g.][]{springel_e_2010}. This is especially relevant near large density contrasts as they can be quickly diluted with mass being pushed from one cell to the next. This is confirmed by this method dispersing the cloud more \textit{quickly} for the higher density contrast and more \textit{slowly} for the lower density contrast, compared to the three methods with moving cells. Though the \textsc{mm} method has a Voronoi mesh which moves with the bulk flow, flux exchanges across cell boundaries are still expected in turbulent regions \citep[e.g.][]{mandelker_thermal_2021}. Since the \textsc{mfm} and \textsc{mfv} methods are similar to the \textsc{mm} method, but have smoothed boundaries, one would expect the \textsc{mm} method to suffer more from numerical mixing and hence to disperse the cloud more quickly. We do not see this reflected in the results for the higher density contrast. The results from the \textsc{mm} method are well matched to the \textsc{mfv} method and cloud dispersal is slightly slower than for the \textsc{mfm} method. 

For both density contrasts, the differences between the \textsc{amr} and \textsc{mm} methods at the highest resolution are surprisingly small. Since the \textsc{mm} method has a moving mesh, one would expect the numerical mixing and hence cloud dispersal to be slower compared to the \textsc{amr} method. It should also be noted that the two methods both seem well converged for this metric.

It is worthwhile to compare the behaviour seen for the higher density contrast with the images in Fig. \ref{fig:slice_plots}, where it can be seen that the morphological evolution of the clouds is very different, even for simulations that look very similar in Fig. \ref{fig:dense_mass}.
We postulate that the slower dispersal of the cloud in the three SPH methods is due to the very sharp and high density contrast, which is hard to resolve for these methods and leads to a much slower decline in cloud mass.
Having the SPH method in the fold for the lower density contrast strengthens our conviction that it is simply unable\footnote{The term "unable" is a misnomer in this context, because in the absence of conduction the formally correct solution is zero mixing.} to handle a high and sharp density contrast. We note that the \textsc{anarchy-pu sph} and \textsc{gdf sph} methods show cloud destruction morphologies more akin to Eulerian methods.

The \textsc{gdf sph} method, whose evolution of dense mass for the higher density contrast is similar to \textsc{Sphenix sph}, gives a much longer dispersal time for the lower density contrast. The difference with the other methods also increases with increasing resolution. This might be due to the many cloudlets produced in this method (see Fig. \ref{fig:slice_plots}) being above $\rho_{\text{cl}}/3$ for the lower density contrast, but below it for the higher density contrast where the initial cloud density is higher. Hence, they would only be included in the lower density contrast's definition of dense gas.  We note, that in line with recent literature \citep[e.g.][]{gronke_growth_2018, gronke_how_2020}, we choose $\rho_{\text{cloud}} / 3$ as our density threshold. \citet[][]{wadsley_gasoline2_2017} (in line with  \citet[][]{agertz_fundamental_2007}) chose a threshold of $0.64 \rho_{\text{cloud}}$, which has a large effect on the evolution of dense gas, resulting in an apparently much faster dispersal of the cloud. We confirmed that using their higher threshold results in the much more rapid drop in the mass of dense mass observed by those authors \footnote{Small changes in the initial conditions can also have a large influence on the morphology of the cloud. We show a comparison between two different initial condition files in Appendix C.}.

We have tested \textsc{sphenix sph} with higher values for the artificial conduction and viscosity parameters. This made cloud dispersal faster by diffusing the edges of the confined cloud to the background density. The morphology remained unchanged and the dense mass evolution was, for the higher density contrast, still distinctly different from the non-SPH methods. In an effort to alleviate any potential problems with neighbour finding in the SPH codes near the edge of the cloud, we also experimented with an increased particle number in the wind while keeping its density fixed, which implies lower-mass wind particles. Even when using the same SPH particle number density for the wind as for the cloud's interior, differences compared to the default set-up were minimal.

Finally, we note some morphological differences evident from Figure \ref{fig:slice_plots}. For the higher density contrast the \textsc{mfm} and \textsc{mfv} simulations create a noticeably smooth density distribution with long wispy tails. The \textsc{amr} simulation shows features associated with a regular grid, being fully symmetric and having a horizontal component aligned with the cloud's original centre. We find the cloud in the \textsc{mm} simulation to be the most disturbed. Even though it retains a denser core of gas, the filamentary tails are much less smooth and show significant substructure. The very different morphologies between the three SPH methods demonstrate the large effect of different artificial viscosity and conduction prescriptions. For the \textsc{Sphenix sph}  simulation the cloud slowly diffuses at the edges without the clear break-up seen in most other methods. The \textsc{gdf sph} simulation does break the cloud, but yields a very clumpy distribution, unlike the smoother tails seen in the \textsc{mfm}, \textsc{mfv}, \textsc{mm} and \textsc{amr} methods. The \textsc{anarchy-pu sph} method does show smoother tails, but also retains a relatively large dense cloud which seemingly only sheds small amounts of mass through its tails without any shattering.



\subsection{Mass of intermediate-temperature gas} \label{sec:results-mass_intermediate_temp}
The mass at intermediate temperatures is defined as

\begin{figure}
	\includegraphics[width=\columnwidth]{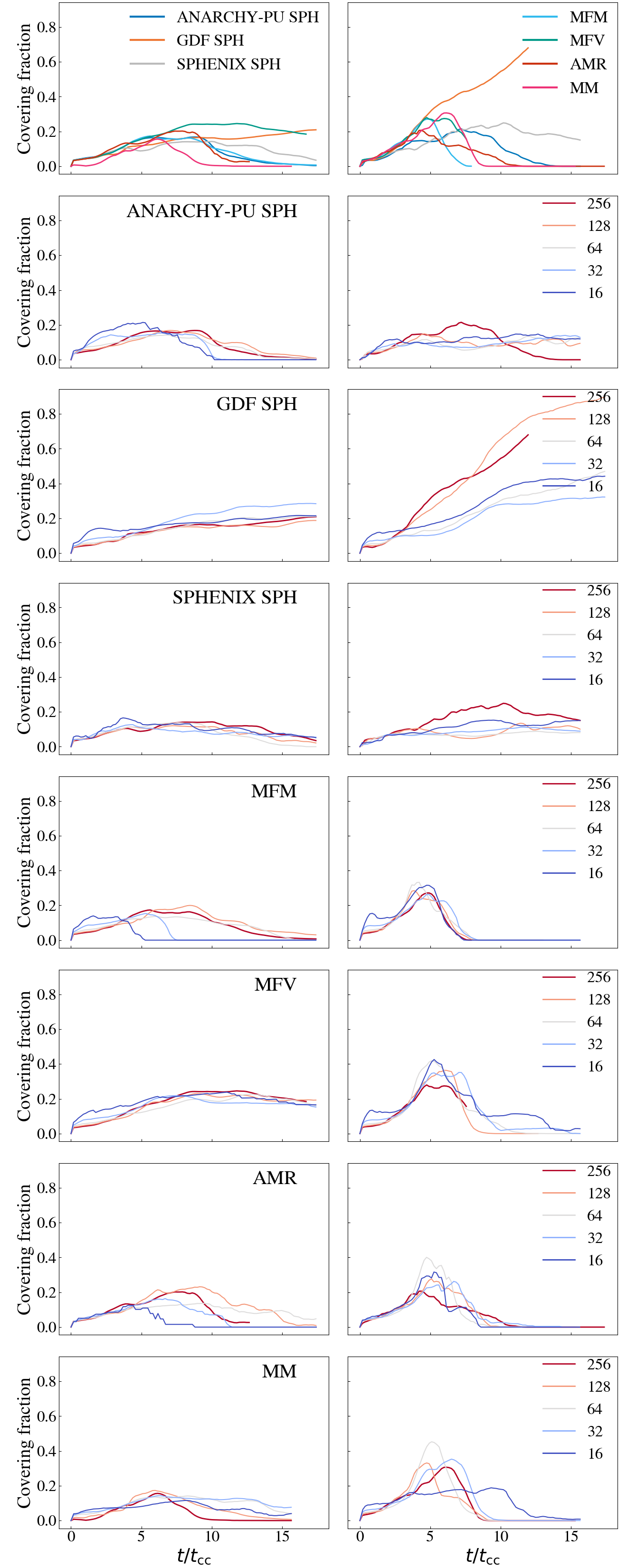}
    \caption{Evolution of the covering fraction of intermediate-temperature gas for different hydro schemes (colours and rows) and different resolutions (linestyles). Initial density contrasts are $\chi= 10$ (\textbf{left}) and $\chi=100$ (\textbf{right})}
    \label{fig:cov_fracs}
\end{figure}

\begin{equation}
    \begin{split}
        M_{\text{mix}} = \sum_i m_i \Bigl[ \log_{10}(T_{\text{mix}}) -& \frac{1}{4}\left(\log_{10}(\chi) \right)\\
        &< \log_{10}(T_i) <\\
        &\log_{10}(T_{\text{mix}}) + \frac{1}{4}\left(\log_{10}(\chi) \right) \Bigr] \, ,
        \label{eq:intermediate_temperature}
    \end{split}
\end{equation}
where $T_i$ is the temperature of a particle or volume element, and $T_{\text{mix}}$ the geometric mean temperature
\begin{equation}
    T_{\text{mix}} = \left(T_{\text{cloud}}T_{\text{wind}}\right)^{1/2} = T_{\text{wind}}\chi^{-1/2} = T_{\text{cloud}}\chi^{1/2} \, .
\end{equation}
This corresponds to the mass within half the logarithmic temperature range between the cold cloud and the hot wind centred on the geometric mean temperature.

\noindent Large differences in the mass of gas with intermediate temperatures would indicate possible large discrepancies in predictions for observable diagnostics of mixing layers.

As our setup is in thermal pressure balance, there is a direct correspondence between the density and the temperature contrasts. Gas with a density $\hat{\chi}$ higher than the initial wind density will have a temperature of $\frac{1}{\hat{\chi}}$ times the initial wind temperature.
For the higher density contrast case studied here $(\chi = 100)$, our intermediate-temperature range yields temperatures that do no overlap with the density range which counted towards the mass of dense gas in \S\ref{sec:results-dense_mass}, however, for the lower density contrast $(\chi = 10)$ the upper half of this range overlaps with the dense gas density range.

In the absence of radiative cooling, this intermediate-temperature gas can be produced by the density changes due to the initial shock, at later times the only mechanisms that can produce this gas phase are numerical mixing and adiabatic expansion. Hence, the peak occurs when the dense gas mass decreases rapidly.

In Figure \ref{fig:intermediate_temp}, similar to Fig. \ref{fig:dense_mass}, we see the dichotomy between the lower and higher density contrasts. For $\chi = 10$ there is a large amount of variation, whereas for $\chi = 100$ all methods are in reasonable agreement except the three SPH simulations. 

For the lower density contrast, the mass of intermediate-temperature gas peaks at roughly the same time for the \textsc{anarchy-pu sph}, \textsc{mfm} and \textsc{amr} methods, but the peak is higher for the  \textsc{amr} method. The \textsc{mm} simulation peaks earlier and the \textsc{mfv} and \textsc{sphenix sph} simulations slightly later. \textsc{gdf sph} is an outlier, showing a slow but persistent increase in the mass of intermediate-temperature gas.

\begin{figure}
	\includegraphics[width=\columnwidth]{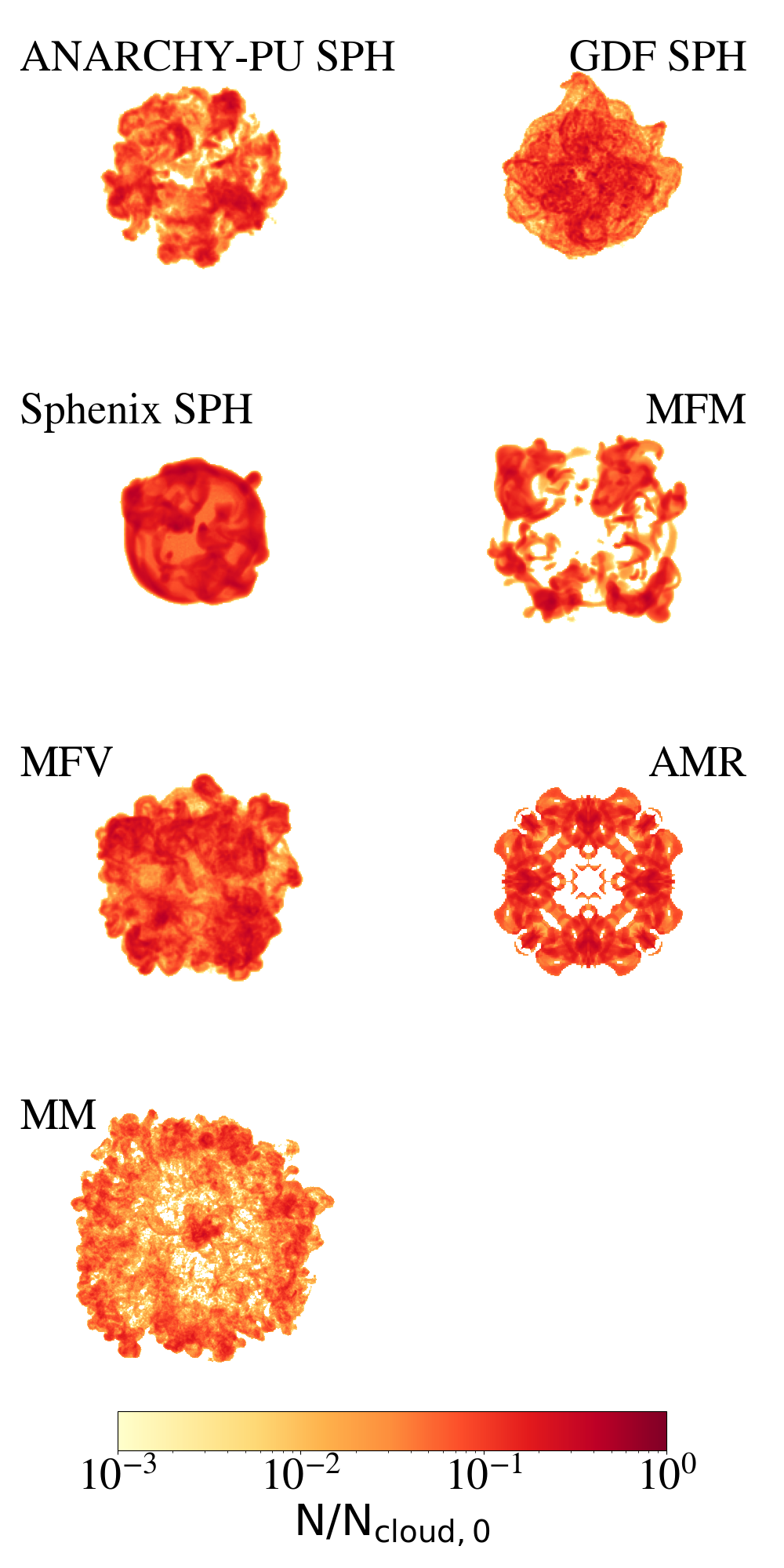}
    \caption{Projection of the mass at intermediate temperatures for initial density contrast $\chi = 10$ at time $t \approx 9\mathrm{t_{\text{cc}}}$. Morphologies differ widely, ranging from fuzzy balls and inky blotches to symmetric grid-like artifacts.}
    \label{fig:cov10}
\end{figure}

For the higher density contrast the four non-SPH methods produce similar results.
The many cloudlets seen in the \textsc{gdf sph} simulations are mostly in this intermediate-temperature regime, giving rise to a large amount of mixed gas, far greater than produced in simulations using any other method. The other two SPH methods match well except at late times when \textsc{anarchy-pu sph} shows a drop in intermediate-temperature gas, whereas the \textsc{Sphenix sph}  method maintains an almost constant level. This is due to \textsc{anarchy-pu sph} having lost most of the cold cloud by this time, ending the creation of new intermediate-temperature gas.

The convergence with resolution is worse than for the dense gas mass. We note that even when the third and fourth resolution level seem converged, the highest level can give very different results, especially at later times. Indicative of the numerical mixing in the \textsc{amr} and \textsc{mm} methods is the decrease in peak height with increasing resolution, especially apparent for the higher density contrast. A higher resolution should give rise to less numerical mixing, and hence allow a sharp density contrast to survive for longer.

For the lower density contrasts differences between the two highest resolutions are very small for the \textsc{gdf sph}, \textsc{Sphenix sph} and \textsc{anarchy-pu sph} simulations, whereas this only hold true for the other methods at early times.

\begin{figure}
	\includegraphics[width=\columnwidth]{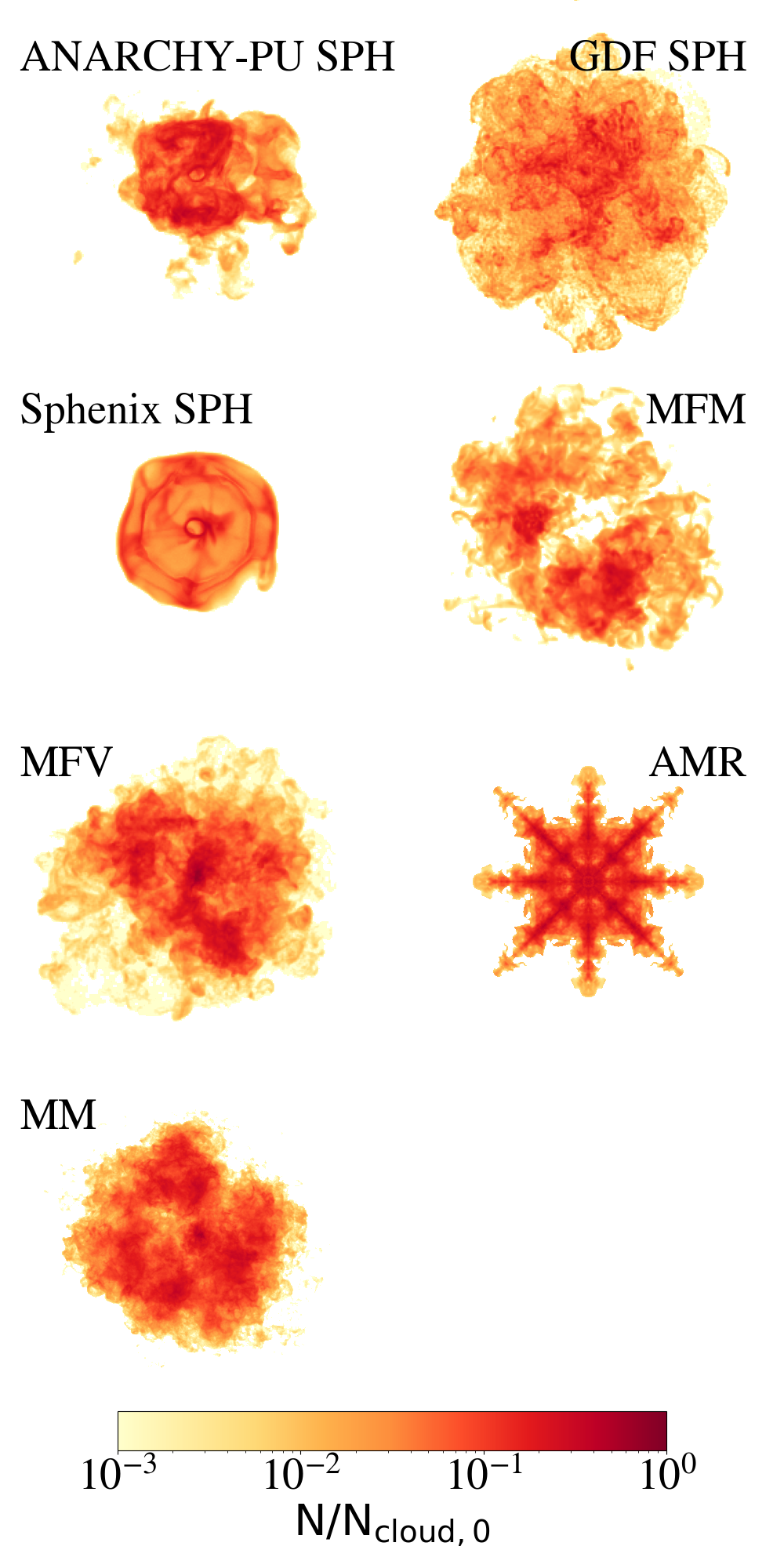}
    \caption{Projection of the mass at intermediate temperatures for initial density contrast $\chi = 100$ at time $t \approx 5\mathrm{t_{\text{cc}}}$. Very different morphologies are apparent, even among those methods for which the covering fractions are similar (e.g. compare \textsc{mm} and \textsc{mfv}). The grid-like artifacts from the \textsc{amr} method are clearly visible.}
    \label{fig:cov100}
\end{figure}

\subsection{Covering fraction of intermediate-temperature gas}
\label{sec:results-covering_fraction}

\begin{figure*}
    \centering
    \includegraphics[width=\linewidth]{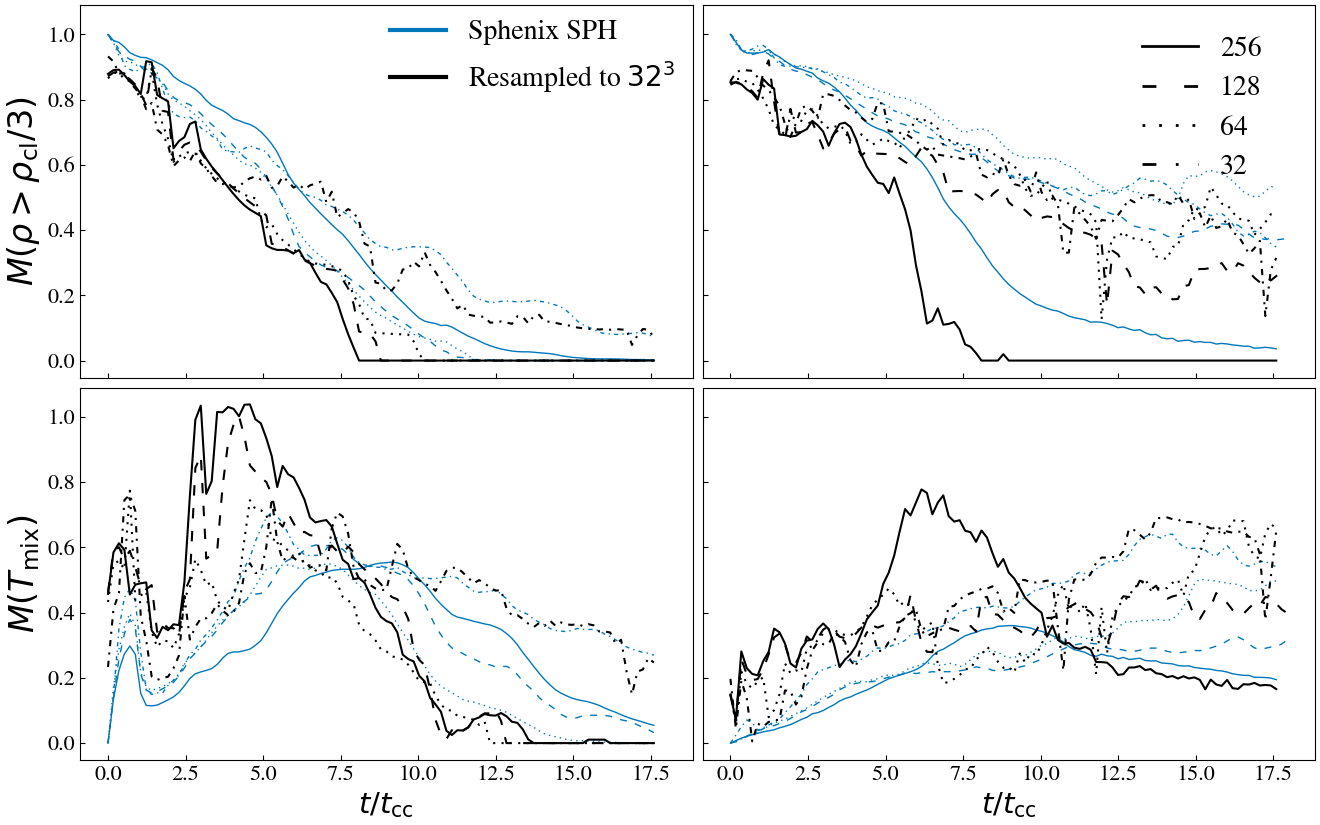}
    \caption{Evolution of the mass of dense gas (top) and intermediate-temperature gas (bottom) for the \textsc{Sphenix sph} hydrodynamics method. Initial density contrasts are $\chi= 10$ (\textbf{left}) and $\chi=100$ (\textbf{right}). The black lines show the evolution of the different gas phases after coarse-graining to the common resolution of $32^3$ grid-cells.}
    \label{fig:sphenix_resampled}
\end{figure*}

\noindent Whereas Figure \ref{fig:intermediate_temp} showed the mass in the intermediate-temperature gas phase, we will now look at the projected distribution of that gas. This is an observationally relevant metric of cloud-wind interactions. Some tentative efforts have already been made to translate the rather idealised cloud-wind interaction to expectations for observable absorption lines \citep[e.g.][]{de_la_cruz_simulated_2020}. Given the large variety of evolutionary paths taken by the cloud depending on the hydrodynamics method, we explore, in a simplified setting, what impact the hydrodynamics solver might have on observable statistics such as absorption lines. To this end, we project all intermediate-temperature gas (defined by eq. \ref{eq:intermediate_temperature}) along the long axis of our simulation box. Counting the fraction of pixels that have a column density of intermediate-temperature gas higher than 1\% of the original cloud column density\footnote{This is an arbitrary choice and does not reflect the sensitivity limits encountered in particular observations.}, we present the results as a function of time, resolution and hydrodynamics method in Fig. \ref{fig:cov_fracs}.

We observe a large variety in covering fractions for the different methods, with very different convergence characteristics. For the higher density contrast, methods which are similar in terms of the evolution of the mass of intermediate temperature gas, namely \textsc{mfm}, \textsc{mfv}, \textsc{mm}, and \textsc{amr}, produce very different covering fractions. Not only does the height of the peak vary, the peaks are also offset by considerable amounts.  

The important metric in this figure is the combination of peak width and height. A higher peak covering fraction indicates a larger likelihood of detecting any material in a hypothetical absorption line study, but for a narrow peak the time-window would be short. We remind the reader that our initial cloud occupies only 3\% of the projected area of the box. With some methods reaching peak covering fractions of well over 50\%, these must be due to low-density fragments as is also evident from Fig. \ref{fig:slice_plots}. 

Representative images of the projected clouds after $9 t_{\text{cc}}$ are shown in Fig. \ref{fig:cov10} for $\chi = 10$ and at $5 t_{\text{cc}}$ in  Fig. \ref{fig:cov100} for $\chi = 100$. The time is chosen to coincide with the peak of the covering fraction evolution. The morphology of the projected intermediate-temperature gas differs significantly. The \textsc{amr} method has very symmetric grid-like characteristics. \textsc{gdf sph} and \textsc{anarchy-pu sph} clearly show that additional methods to reduce artificial surface tension produce more diffuse projected distributions compared to \textsc{Sphenix sph}. \textsc{mfv} and \textsc{mm} are quite comparable in the total extent of their gas distribution, with \textsc{mfm} showing more clumping, and for $\chi = 100$ \textsc{mm} retaining a denser core. Except for the \textsc{Sphenix sph} and \textsc{amr} methods, none of the projected distributions are symmetric. The \textsc{amr} method shows a grid symmetry, and \textsc{Sphenix sph} shows the effects of artificial surface tension retaining a near perfect spherical symmetry, but none of the other methods shows either of these symmetries. Given that the initial conditions are set up on a grid, symmetry is expected. We tried to explicitly enforce symmetry by only generating half of the initial conditions and creating the other half by mirroring the first, but this did not lead to appreciable more symmetric morphologies.


\begin{figure*}
    \centering
    \includegraphics[width=\linewidth]{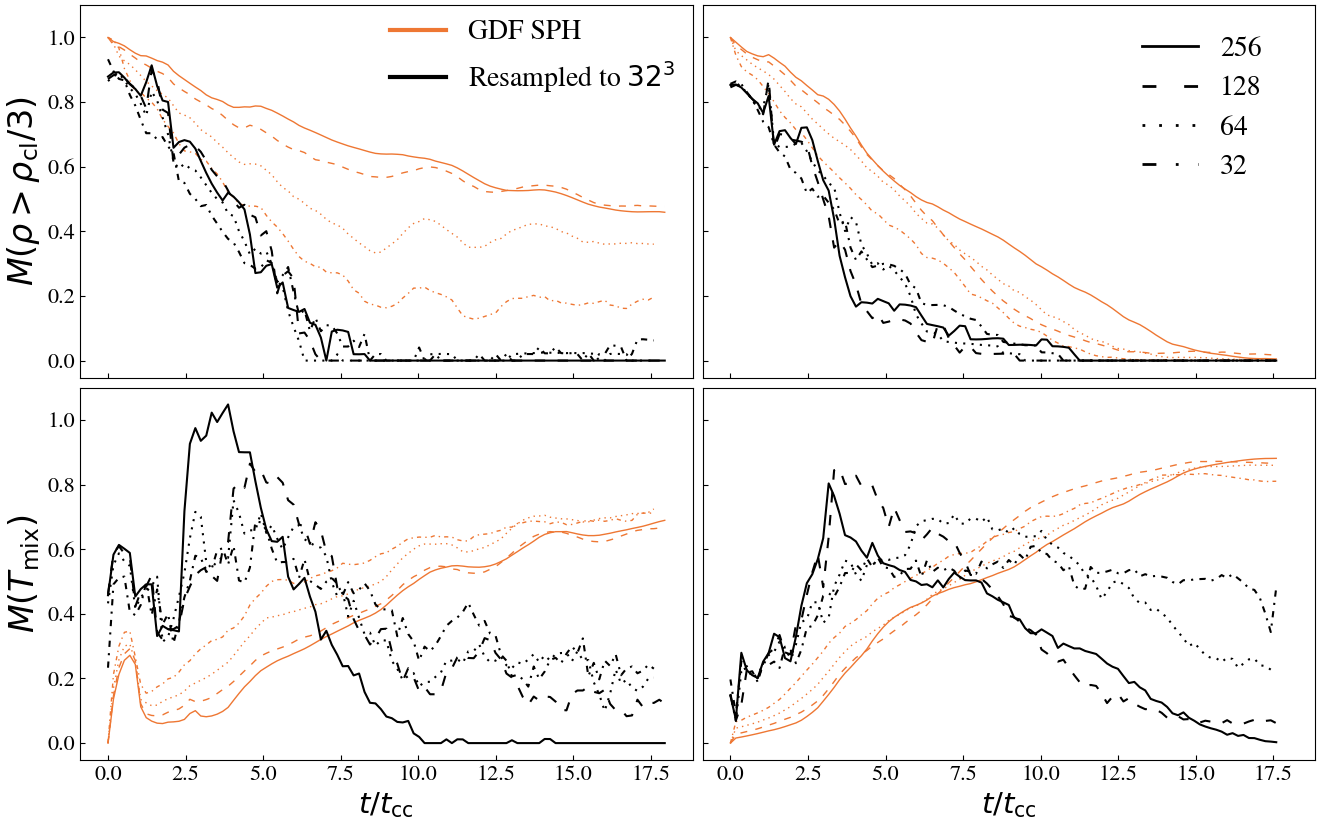}
    \caption{Evolution of the mass of dense gas (top) and intermediate-temperature gas (bottom) for the \textsc{gdf sph} hydrodynamics method. Initial density contrasts are $\chi= 10$ (\textbf{left}) and $\chi=100$ (\textbf{right}). The black lines show the evolution of the different gas phases after coarse-graining to the common resolution of $32^3$ grid-cells. When coarse-grained, the many cloudlets produced by \textsc{gdf sph} contributing to both the dense gas (at their cores) and intermediate-temperature gas (at their edges) average out with the hot wind, leading to an evolution more comparable to that predicted by other methods.}
    \label{fig:gasoline_resampled}
\end{figure*}

\subsection{Coarse-graining to a common resolution} \label{sec:downsampling}

A new avenue in exploring the evolution of idealised hydrodynamics simulations is to compare the level of mixing in lower-resolution simulations to that in higher-resolution simulations after coarse-graining the latter onto a lower-resolution grid. To this end we have devised a 3D grid which is comoving with the median position of the dense gas in a simulation. On this 3D grid, we compute mass-weighted quantities for a grid resolution corresponding to the one but lowest resolution employed in this work, $32^3$ grid cells. No smoothing is applied in this procedure, volume elements whose centres fall within a grid cell are fully counted in that single grid cell.

We find that in general, the coarse-grained results differ from the corresponding lower-resolution simulation, though they are more similar to them than the non-coarse-grained higher-resolution simulations. As a representative comparison with coarse-grained results, we show in Fig. \ref{fig:sphenix_resampled} a comparison between \textsc{Sphenix sph} at all resolutions and when coarse-grained onto a regular grid with $32^3$ grid cells. Though there are clear differences between the coarse-grained and original resolution results, they agree qualitatively. For the higher density contrast ($\chi = 100$), we find that there is a consistent offset between the converged full resolution simulation and the coarse-grained results, which show similar convergence. The coarse-grained mass of dense gas is slightly lower at all times, whereas the coarse-grained mass of intermediate-temperature gas is higher compared to the full resolution simulations. This is in line with expectations, where dense gas averaged over a finite volume might drop below the density threshold, and likewise hot and cold gas numerically mix on the coarse-grained grid resolution, yielding more intermediate-temperature gas. Similar small deviations are found for all but the \textsc{gdf sph} hydrodynamics method.

For \textsc{gdf sph} we find differences which change the interpretation of the results. As shown in Fig. \ref{fig:gasoline_resampled}, for the lower density contrast ($\chi = 10$), the simulation employing this hydrodynamics method retain a large amount of dense gas at the highest resolution, but after coarse-graining this onto the lower-resolution grid, the dispersal time is in line with the other hydrodynamics methods. We attribute this to the large amount of tiny cloudlets formed using this method (see Fig. \ref{fig:slice_plots}). When averaging over a grid cell containing a cloudlet and dilute gas, the average density will be below the dense gas threshold. 

For intermediate-temperature gas it is not just the time scale which changes, but also the shape of the curve. Whereas the original \textsc{gdf sph} shows a continuous increase of intermediate-temperature gas with time, unlike the peaked time evolution seen for other methods, after coarse-graining peaks do appear (see Fig. \ref{fig:gasoline_resampled}). The time at which the amount of intermediate-temperature gas peaks now also coincides with the time at which dense gas is disappearing the fastest. Just as for the dense mass, we attribute this to the survival of many small cloudlets. These fall into the intermediate-temperature regime at late times, but when averaged over a finite volume, their temperature no longer satisfies the criterion for intermediate-temperature gas.

These coarse-grained evolutionary tracks would be important for hypothetical observations measuring volume-averaged densities or temperatures. However, because the ionisation balance and molecule abundances depend on the resolved density and temperature, it is the original non-coarse-grained result that is most relevant for observations. This is an important distinction when making predictions for observational diagnostics, since we have shown that the two approaches can yield qualitatively different results.

\section{Conclusions} \label{sec:conclusions}
We conducted a systematic study of the non-radiative 3D cloud-wind problem using seven different hydrodynamics solvers. Three SPH methods: \textsc{sphenix} \textsc{sph} , \textsc{gdf sph}, \textsc{anarchy-pu} \textsc{sph}; two meshless methods: meshless finite mass (\textsc{mfm}), and meshless finite volume (\textsc{mfv}); a static grid with adaptive mesh refinement (\textsc{amr}); and a moving mesh (\textsc{mm}). For all methods we conducted simulations for two initial density contrasts: $\chi = \{10, 100\}$, and four resolutions: $nR_{\text{cloud}} = \{1.6, 3.2, 6.4, 12.8\}$ wind particles per cloud radius. We simulate a static spherical cloud that is initially in pressure equilibrium and is impacted by a wind with Mach number $\mathcal{M}=1.5$. As we do not include physical conduction, mixing is numerical. However, because pressure equilibrium is broken due to the ram pressure and the initial shock from the wind, a short entropy generating phase is present breaking pressure equilibrium. At later times the densities and temperatures can then evolve adiabatically. 

We compared the mass of dense gas and the mass of intermediate-temperature gas as a function of time, as well as the covering fraction of the intermediate-temperature gas.  We find that the differences between hydrodynamics methods are substantial, and are highly dependent on the quantity of interest.
Summarised, we find:
\begin{enumerate}
    \item For $\chi = 100$ the differences are much more pronounced compared to $\chi = 10$. For the lower density contrast most methods stretch the cloud into a thin shell before fragmenting it, and all tails are relatively short (Fig. \ref{fig:slice_plots}).
    For $\chi = 100$ in the simulations using the \textsc{mfm}, \textsc{mfv}, \textsc{mm} and \textsc{amr} methods the cloud fragments after an initial compression, but the resulting morphologies differ widely. 
    The \textsc{mfm} and \textsc{mfv} methods predict tails that are longer than for \textsc{mm} and \textsc{amr}. The three SPH methods show a very different evolution, with a more pronounced core of dense gas at $t = 5t_{\mathrm{cc}}$. For \textsc{Sphenix sph} the cloud survives as a single entity, only losing mass through slow diffusion. For \textsc{gdf sph} the cloud fragments into many cloudlets that experience only limited further evolution.
    For \textsc{anarchy-pu sph} the cloud ablates into a smooth tail similar to the four non-SPH methods. 
    
    \item The evolution of the mass of dense gas (Fig. \ref{fig:dense_mass}) differs greatly with the initial density contrast and hydrodynamics method. For $\chi = 10$ the mass of dense gas is reduced by a factor of 7 within $\approx 10$ cloud crushing times, except for \textsc{gdf sph} which predicts a much slower decline. At the highest resolution level, \textsc{Sphenix sph} and \textsc{mfv} also deviate slightly at all times from the other four methods which seem well converged. The cloud lifetime increases with resolution. All methods converge at the one-but-highest resolution level, \textsc{Sphenix sph} already converges at lower resolution. The two meshless methods (\textsc{mfm} and \textsc{mfv}) convergence most slowly. For the higher initial density contrast ($\chi = 100$) the differences in lifetime are even more pronounced. For \textsc{mfm, mfv, amr} and \textsc{mm} the lifetime is half that for the lower density contrast, and the convergence is better.  The convergence of the \textsc{sph} methods becomes worse compared with the lower density contrast, especially at late times. However, whereas the \textsc{gdf sph} lifetime decreases with density contrast, the lifetime seems constant for \textsc{Sphenix sph} and increases for \textsc{anarchy-pu sph}.
    
    \item The mass of intermediate-temperatures gas peaks at the time when the cloud loses mass most rapidly (Fig. \ref{fig:intermediate_temp}). As the time of destruction differs between the methods, the peak in this quantity occurs at different times. For the higher density contrast ($\chi = 100$) the height of the peak is similar for \textsc{mfm}, \textsc{mfv}, \textsc{mm} and \textsc{amr}, but significantly lower for \textsc{Sphenix sph} and \textsc{anarchy-pu sph} due to the tardy disappearance of the cloud in those methods. \textsc{gdf sph} shows a continuous rise in the mass of intermediate-temperature gas, which is ascribed to the fragmentation into many long-lived cloudlets. However, after coarse-graining the higher-resolution \textsc{gdf sph} simulations to a lower common resolution, a peak in the mass of intermediate-temperature gas becomes apparent (Fig. \ref{fig:gasoline_resampled}), which is more in line with the other hydrodynamics methods.
    For $\chi=10$ the longer survival of dense gas for the \textsc{mfv} method translates into a significant offset in this metric. The \textsc{amr} method produces more intermediate-temperature gas than any other method, which might be explained by numerical mixing in grid-based simulations. Whereas \textsc{amr} and \textsc{mm} converged to the same answer for the dense gas, they are very different in this metric, with \textsc{mm} producing less intermediate-temperature gas and for a shorter time.
    
    \item The evolution of the covering fraction of intermediate-temperature gas, a metric relevant for the probability of observing intermediate-ionisation stages in absorption, differs greatly both in terms of the time and width of the peak (Fig. \ref{fig:cov_fracs}). Methods which yield similar masses of intermediate-temperature gas can have very different covering fractions. For the lower density contrast 
    the covering fractions decline relatively slowly with time, compared to the much faster changes in the mass contributing to this covering fraction. \textsc{mfv} and \textsc{gdf sph} are again the outliers, predicting a much slower decline. For the higher density contrast, the peaks in the covering fraction are more pronounced, but peak heights differ by more than 50\%, something that is not visible in the mass of intermediate-temperature gas.
    The covering fraction peaks after the time when most intermediate-temperature gas is present. The three \textsc{sph} methods, particularly \textsc{gdf sph}, predict much larger covering fractions at late times.
    For the lower density contrast all methods except \textsc{mm} and \textsc{amr} are well converged at the one-but-highest resolution level, with \textsc{Sphenix sph} and \textsc{mfv} showing convergence at lower resolution. The convergence is generally better for the higher density contrast, except for \textsc{gdf sph} and \textsc{mm}.
    

\end{enumerate}
Fifteen years after \citet[][]{agertz_fundamental_2007} provided the comparison between an SPH and a grid code for the cloud-wind interaction, we conclude that at a lower density contrast ($\chi = 10$) most methods now agree well on the evolution of dense gas, though at a higher density contrast ($\chi = 100$) significant differences remain, also between non-SPH methods. We introduced new metrics to study the properties of mixed intermediate-temperature gas and concluded that for those quantities predictions from cloud-wind interaction studies are still marred by numerical uncertainties. 
Even among the three methods which yield reasonably similar results, i.e. \textsc{mfm}, \textsc{mm} and \textsc{amr}, the relevant timescales differ by a factor two. For $\chi = 10$ \textsc{mfv}, \textsc{Sphenix sph}, and particularly \textsc{gdf sph} \footnote{There is significant disagreement with the results in \citet{wadsley_gasoline2_2017}, with our results showing more dense clumps and a slower decrease in the dense gas mass. However, the degree and origin of those differences is unclear.}, predict a slower dispersal of the dense gas, and a corresponding later peak in the mass of intermediate-temperature gas. For $\chi = 100$ all \textsc{sph} methods show a slower dispersal, though they also differ strongly from each other. The covering fractions of intermediate-temperature gas differ even more, with all methods showing different peak times and evolution. Hence, we conclude that the effects of the inclusion of additional physics, such as radiative cooling, MHD or physical conduction, should be studied by comparing simulations using both the same and different hydrodynamics methods.
Ideally, different hydrodynamics methods would be used for any cloud-wind interaction study to both quantify uncertainties and offer greater understanding.

\section*{Acknowledgements}
The authors thank Bert Vandenbroucke for technical help in realising this project, and Mike Shull for helpful comments. We are also grateful to the anonymous referee for their helpful comments. The authors gratefully acknowledge funding from the Netherlands
Organization for Scientific Research (NWO) through Vici grant
number 639.043.409 and research programme Athena 184.034.002 from the Dutch Research Council (NWO). This work used the DiRAC@Durham facility managed by the Institute for Computational Cosmology on behalf of the STFC DiRAC HPC Facility (www.dirac.ac.uk). The equipment was funded by BEIS capital funding via STFC capital grants ST/K00042X/1, ST/P002293/1, ST/R002371/1 and ST/S002502/1, Durham University and STFC operations grant ST/R000832/1. DiRAC is part of the National e-Infrastructure.
The research in this paper made use of the SWIFT open-source simulation code
(http://www.swiftsim.com, \citep[][]{schaller_swift_2018}) version 0.9.0. This paper made use of the following \textsc{python} packages: \textsc{swiftsimio} \citep[][]{borrow_swiftsimio_2020}, \textsc{numpy} \citep[][]{harris_array_2020}, \textsc{yt} \citep[][]{turk_yt_2011}, and \textsc{matplotlib} \citep[][]{hunter_matplotlib_2007}.

\section*{Data Availability}
All code and initial conditions used to generate the simulations within \textsc{swift} are made part of \textsc{swift}. As the simulations presented in this paper are relatively small simulations that can easily be repeated using the publicly available initial conditions, the data is not made immediately available.



\bibliographystyle{mnras}
\bibliography{main_paper} 




\appendix
\section{Late time evolution}
\begin{figure*}
    \centering
    \includegraphics[width=\textwidth]{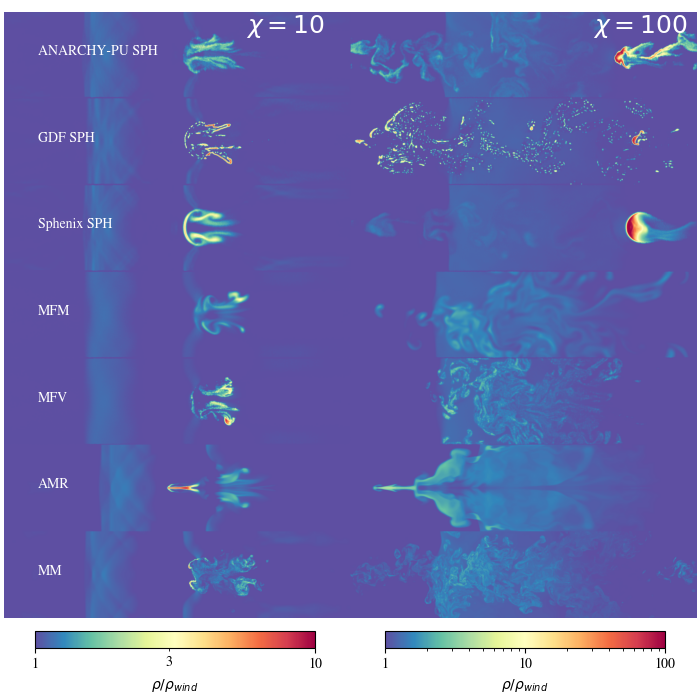}
    \caption{Density slices of the simulation box at $t \approx 10t_{\text{cc}}$, for initial density contrasts $\chi = 10$ (left) and $\chi = 100$ (right), at a resolution of $n=128^3$. Large morphological differences can be observed.}
    \label{fig:projection_plots_late}
\end{figure*}

At late times the solutions of different solvers look far more similar, especially for the lower initial density contrast. Fig. \ref{fig:projection_plots_late} shows the morphology of the clouds at a much later time (10$t_{cc}$) compared to Fig. \ref{fig:slice_plots} (5$t_{cc}$). Though for the higher density contrast the SPH solvers differ from the other methods in that they keep the cloud together, at the lower density contrast everything looks very similar.

\section{Entropy}
\begin{figure*}
    \centering
    \includegraphics[width=\textwidth]{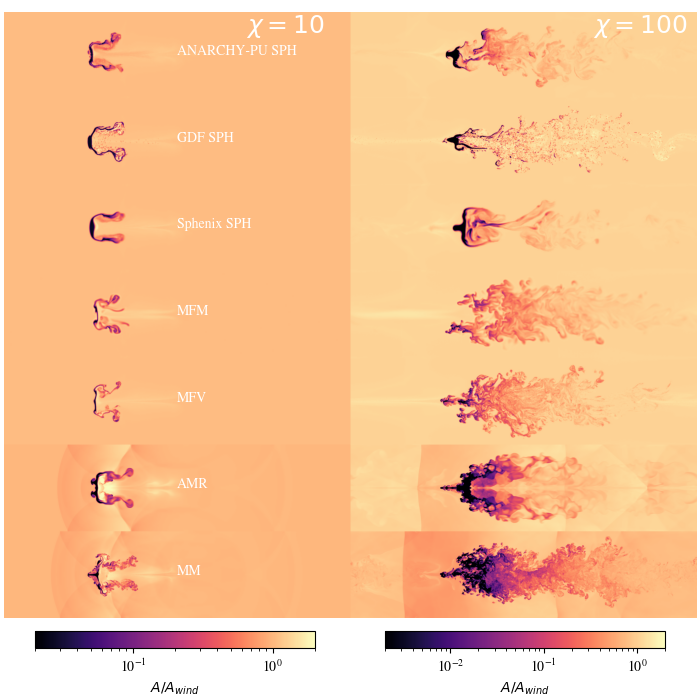}
    \caption{Entropy slices of the simulation box at $t \approx 5t_{\text{cc}}$, for initial density contrasts $\chi = 10$ (left) and $\chi = 100$ (right), at a resolution of $n=256^3$. Large morphological differences can be observed.}
    \label{fig:projection_plots_entropy}
\end{figure*}

Fig. \ref{fig:projection_plots_entropy} shows infinitesimally thin slices of the hydrodynamic entropy 
\begin{equation}
    A = T / \rho^{2/3} \, ,
\end{equation}
with $T$ the temperature and $\rho$ the density of each volume element. We note that these projections, especially for the lower density contrast, look far more alike than the density projections. For the higher density contrast, the differences are larger, but the methods still look much more alike than for their density projections. We note that some authors show entropy projections, instead of densities. Seeing the differences between the two in our work, we urge caution in interpreting differences between methods when only one of these types of images is available for each method.

\section{The impact of different initial conditions} \label{sec:ic_comparison}
\begin{figure*}
    \centering

    \centering
    \includegraphics[width = \textwidth]{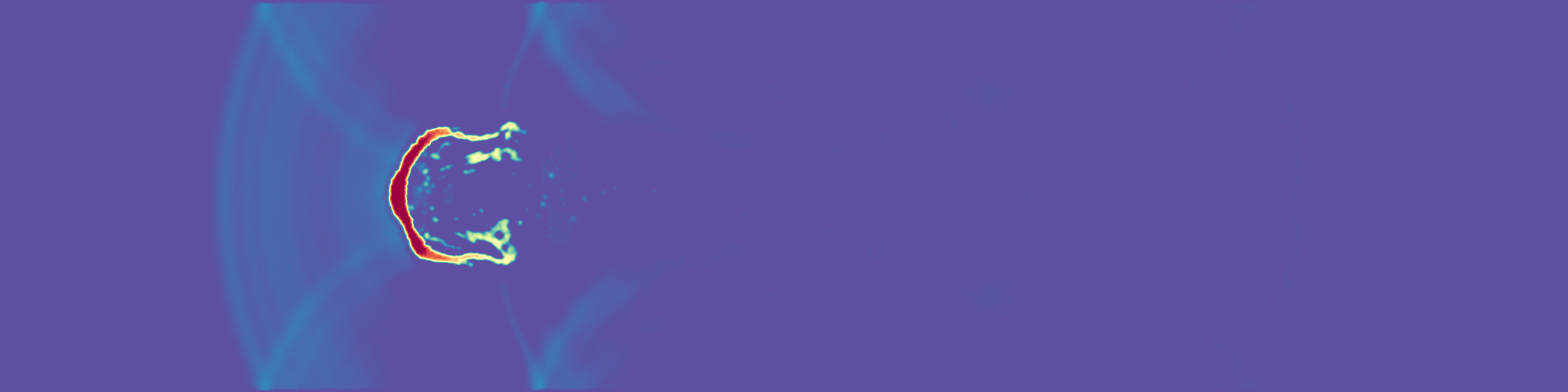}

    \centering
    \includegraphics[width = \textwidth]{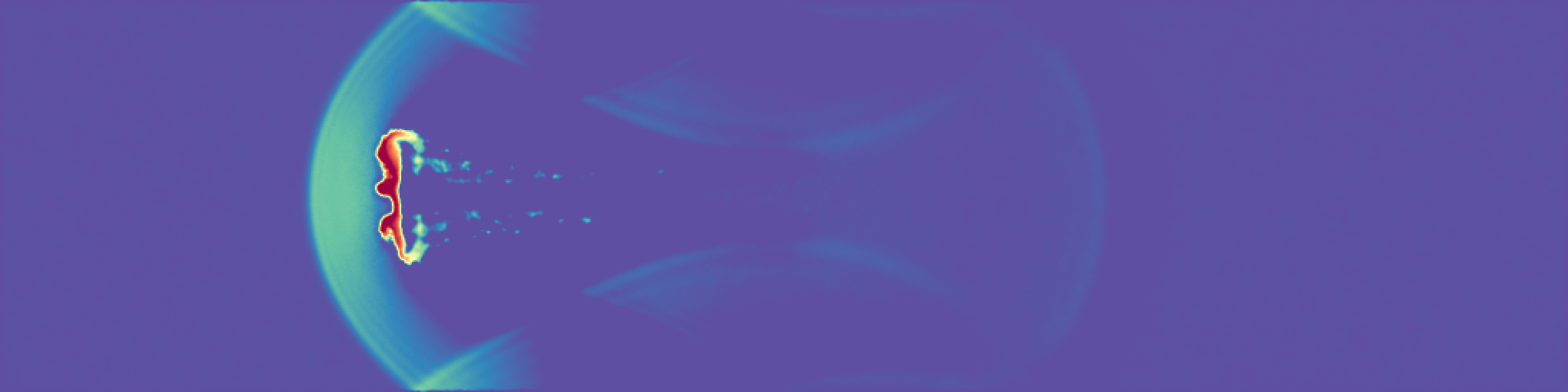}
    \caption{Sliced density at $t \approx 5t_{\text{cc}}$ for a cloud-wind interaction using our ICs (top) and ICs obtained from one of the authors of Wadsley et al. 2017 (bottom), both using the \textsc{gdf sph} implementation in SWIFT.}
    \label{fig:ic_comparison}
\end{figure*}

The choice of initial conditions can have a large impact on the evolution of the clouds. In this study we used a uniform set of initial conditions for all methods. To show what influence different initial conditions can have, we used those from \citet[][]{wadsley_gasoline2_2017} (private communication) and performed a simulation using the same parameters as we have employed in this work. Fig. \ref{fig:ic_comparison} shows a projection of the density using our ICs (top), or those from 
\citet[][]{wadsley_gasoline2_2017}, at the same number of cloud crushing times. Both IC files have a density contrast of $\chi = 10$, and are run with the same \textsc{gdf sph} code (implemented in \textsc{swift}).

Clearly, small changes in the ICs can cause large scale morphological differences between the clouds.

\begin{figure*}
    \centering

    \centering
    \includegraphics[width = \textwidth]{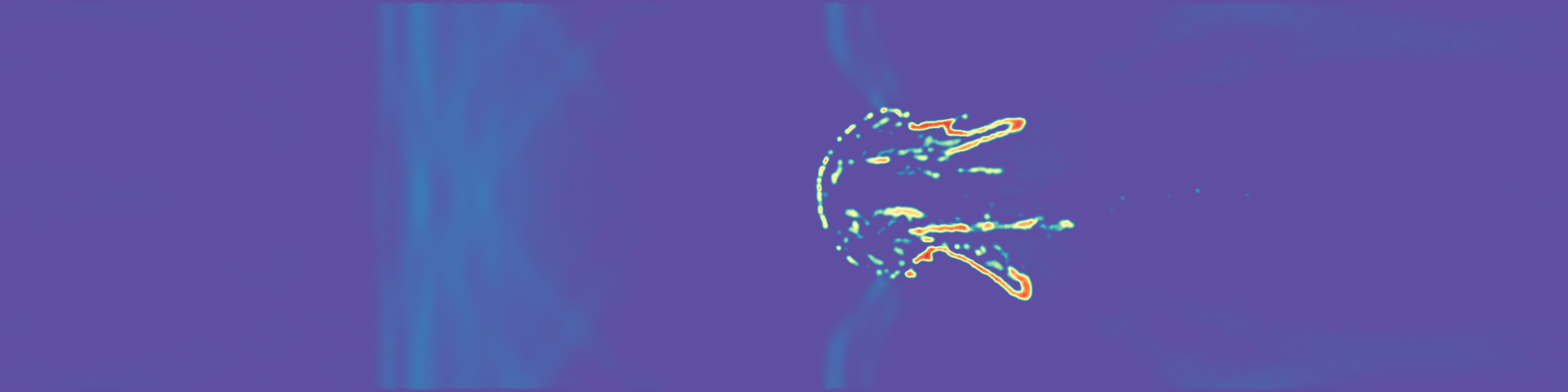}

    \centering
    \includegraphics[width = \textwidth]{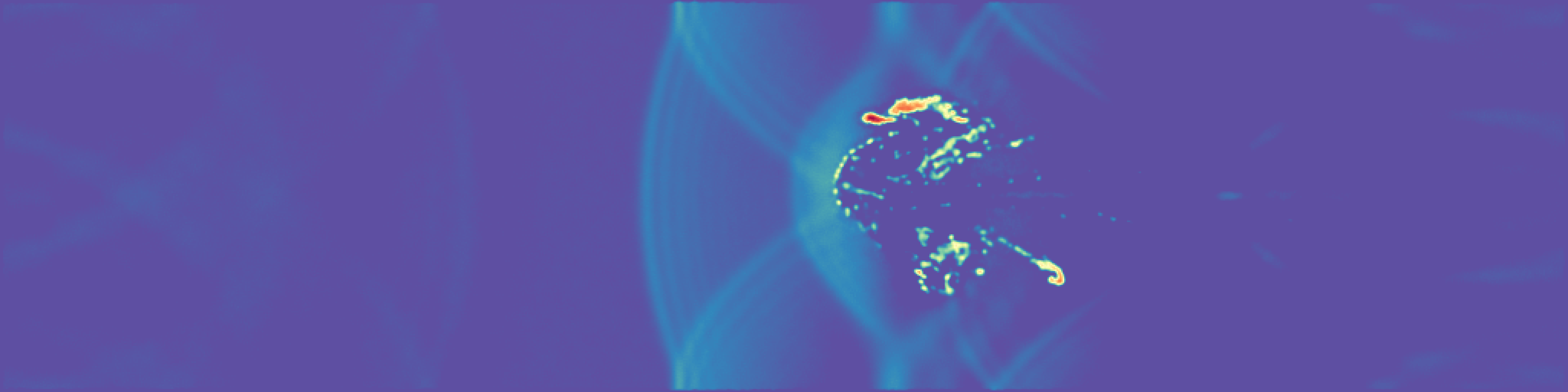}
    \caption{As Figure \ref{fig:ic_comparison}, but at the later time $t \approx 10t_{\text{cc}}$.}
    \label{fig:ic_comparison_2}
\end{figure*}

\bsp	
\label{lastpage}
\end{document}